\documentclass[twocolumn,showpacs,superscriptaddress,10pt,pra]{revtex4-1} 
\usepackage{amsmath,amssymb,graphicx,physics,float}
\usepackage{xcolor}

\begin{document}

\title{Interference-induced directional emission from an unpolarized two level emitter into a circulating cavity}
\author{Lucas Ostrowski}
\author{Scott Parkins}
\affiliation{Dodd-Walls Centre for Photonic and Quantum Technologies, New Zealand}
\affiliation{Department of Physics, University of Auckland, New Zealand}
\author{Morito Shirane}
\author{Mark Sadgrove}
\affiliation{Department of Physics, Faculty of Science, Tokyo University of Science, 1-3 Kagurazaka, Shinjuku-ku, Tokyo 162-8601, Japan}

\begin{abstract}
Chiral coupling between quantum emitters and evanescent fields allows directional emission into nanophotonic devices and is now considered to be a vital ingredient for the realization of quantum networks. However, such coupling requires a well defined circular dipole moment for the emitter -- something difficult to achieve for solid state emitters at room temperature due to thermal population of available spin states. Here, we demonstrate that a two level emitter with a randomly polarized dipole moment can be made to emit directionally into a circulating cavity if a separate emitter is chirally coupled to the same cavity, for the case when both emitter-cavity couplings are strong but in the bad-cavity regime. Our analysis of this system first considers a transient scenario, which highlights the physical mechanism giving rise to the directional emission of the two level emitter into the cavity. An alternative setup involving a weak laser field continuously driving the system is also considered, where the directionality (our proposed figure of merit for this scheme) is shown to be significantly more robust against noise processes. The results presented here take the form of approximate analytical expressions backed by complete numerical simulations of the system.
\end{abstract}

\maketitle

\section{Introduction}
For a number of years it has been recognized that single quantum emitters (QEs) coupled strongly  to an optical cavity are excellent candidates for nodes in quantum networks where matter qubits based on QE are interfaced with flying qubits~\cite{KimbleQI}. In particular, protocols exist for the direct quantum state transfer of qubits from light to atoms and back using such nodes~\cite{Boozer,Dayan1,Dayan2}. However, one problem with such cavity based nodes is that the inherent left-right symmetry of the cavity means that photons couple randomly to the left and right propagating cavity modes, meaning that any state transfer protocol is non-deterministic. Although use of a one-sided cavity solves this problem after a fashion, it also inconveniently restricts the topology of the network.

Recently, the rise of nanophotonic devices and associated evanescent light-matter coupling schemes has led to an elegant work around for the above problem, which goes as follows: While the \emph{total} spin angular momentum of the evanescent field is typically zero, \emph{locally} it is elliptically polarized and, crucially, this local elliptical polarization \emph{is coupled to the direction of propagation of the waveguide mode.} This property is variously referred to as spin-orbit coupling of light~\cite{Liberman,Nori1}, and spin-momentum locking~\cite{Jacob,Bliokh}. This fact means that quantum emitters which support a circularly polarized dipole transition (e.g., a magnetic sublevel) with a suitably oriented quantization axis will couple strongly only to the waveguide mode whose local polarization at the position of the emitter matches the transition's polarization, and thus (due to the coupling of local polarization state and mode propagation direction) couple strongly to only one direction.
The research field which studies light-matter interactions with spin-momentum-locked light is known as ``chiral quantum optics"~\cite{ChiralQO}. In the past few years, a number of devices have been proposed and demonstrated using the principles 
of chiral quantum optics~\cite{Turchette, RauschenbeutelCirc, DayanRefl,RauschIsolator,Shomroni, Sollner}, and chiral techniques are now expected to be a novel tool for the realization of quantum networks~\cite{Pichler,Stannigel}.

\begin{figure}[htb]
	\centering
	\includegraphics[width=0.95\linewidth]{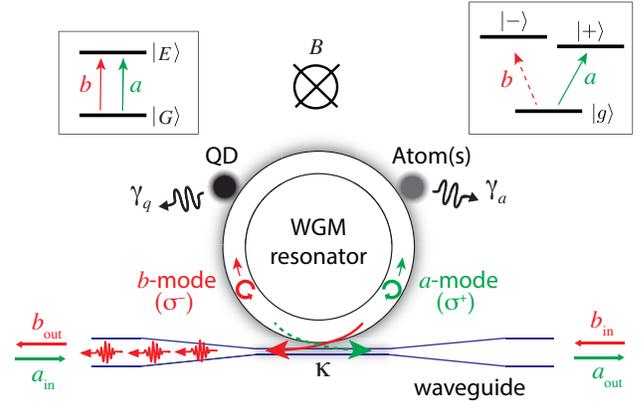}
	\caption[]{Illustration of the system under consideration. A two-level quantum dot (QD) and a V-type three-level atom (or atoms) are coupled to the same whispering-gallery-mode (WGM) resonator.  We label the counter-clockwise circulating mode (polarization $\sigma^+$) of the resonator $a$ and the clockwise circulating mode (polarization $\sigma^-$) $b$. The depicted decay channels are atomic spontaneous emission at rate $\gamma_a$, quantum dot spontaneous emission at rate $\gamma_q$, and decay from each cavity mode at rate $\kappa$ into a waveguide (which could, e.g., be a tapered optical fiber). A magnetic field $B$ may also be applied as shown. }
	\label{fig:Concept}
\end{figure}

One practical problem when it comes to applying chiral coupling techniques is that room temperature solid state quantum emitters typically emit randomly polarized photons~\cite{Abe}. In this case directional emission of photons using the spin-momentum locking effect is not possible. Here, we will demonstrate how it is possible to arrange directional emission from such an unpolarized quantum emitter into a circulating cavity mode if a separate emitter is chirally coupled to the same cavity. Our scheme works as follows: First, assume that a two level quantum emitter -- which we will refer to as a quantum dot (QD) -- 
couples equally, with strength $g_q$, to both counter-clockwise- and clockwise-propagating modes (henceforth referred to as modes $a$ and $b$ respectively) of a whispering-gallery-mode (WGM) cavity. 
Additionally, we assume that the coupling is in the so-called bad-cavity regime, i.e., $g_q\ll\kappa$, with $\kappa$ the cavity field decay rate, but $g_q$ is still assumed to be much larger than the excited state spontaneous emission rate $\gamma_q$ of the QD, so that the cooperativity $C_q=g_q^2/(\kappa\gamma_q)\gg 1$. This ensures 
that any emission from the quantum dot takes place primarily through the cavity modes. Now suppose that another quantum emitter -- which we shall  refer to as an atom -- is prepared in a (stable) internal state which couples chirally to the same cavity, i.e., it has a large coupling strength $g_{a}$ to mode $a$, but a much smaller coupling strength $g_{b}$ to mode $b$.
Later, we will consider the concrete example of a $^{133}$Cs atom, for which the ratio $g_{a}/g_{b}$ can be as large as $\sqrt{45}$. 
Now, if the atom-cavity coupling is also in the bad-cavity regime and with large cooperativity ($C_a=g_a^2/(\kappa\gamma_a)\gg 1$), we will show that the QD excited state decays principally into just one mode (i.e., one direction) as a result of destructive quantum interference between the QD and atomic dipole fields in one polarization. The direction can be chosen by the atomic internal state, giving all of the advantages of chiral coupling even though the quantum dot is randomly polarized.

We note that an alternative scheme also exists where the atom is in the strong coupling regime, i.e., $g_a\gg\kappa,\gamma_a$; here, the mechanism is the strong vacuum Rabi-splitting of one cavity mode, which simply drives that mode off resonance from the QD, leading to Purcell enhancement for QD emission just into the other cavity mode. However, the bad cavity condition is less strict than the strong coupling regime, and, indeed, can be achieved even for very lossy resonators such as plasmonic nanostructures. We therefore focus on the scheme in which both quantum dot and atom are coupled to the cavity in the bad cavity regime.

At this point, we address one obvious question regarding our work: if we assume the existence of a chirally coupled 
atom, then why not dispense with the quantum dot and simply use the chiral coupled photons from the atom itself?
There are several reasons why our scheme has value in spite of this obvious objection. First, although for simplicity we consider a single chirally coupled atom in the present work, there is actually no need for a \textit{single} chirally coupled emitter - our scheme also works with collectively coupled chiral emitters and indeed there are some advantages in terms of a reduction in the required single emitter coupling strength in this case~\cite{ScottMaarten}. Moreover, it is in general much easier experimentally to couple multiple atoms to a resonator, and it can feasibly be achieved for room temperature atoms as opposed to laser-cooled ones~\cite{Ritter}. In this case,  our scheme has the advantage that the achievement of a single emitter is only necessary in the experimentally simpler case of the quantum dot or other solid state emitter.

Second,  the chirally coupled emitter(s) can in principle control  the direction of emission from a number of quantum dots coupled to the cavity, assuming that the quantum dots are addressed one at a time. 

Third, it is useful both conceptually, and potentially in applications, to separate the functions of chiral coupling and single photon emission. This is in line with the general principle of hybrid quantum systems, wherein the advantages of individual quantum systems are hybridized by coupling them. In this case, the convenience of single photon emission from a fixed, single quantum dot and the directionality of coupling for cold atoms are combined to create a single system more useful than either of the systems alone. 

The rest of the paper proceeds as follows: In Section~\ref{SecModel} we introduce our formal model of the system,
and define the relevant master equation. In Section~\ref{SecSingleEx} we  analyse the single excitation regime, where the QD starts in the excited state, the atom is in its ground state, and both cavity modes are in the vacuum state. We perform a trajectory analysis and derive the directionality of emission in the ideal case. In Section~\ref{SecWeaklyDriven}, we consider the case where the quantum dot is weakly driven by an external field. We derive the steady state system properties, and calculate the directionality in the steady state limit.  
The photon statistics of the cavity output fields are also examined in detail. Finally in Section~\ref{SecDisc}, we discuss our results and offer conclusory remarks.

\section{Theoretical model}
\label{SecModel}

\subsection{Master Equation}
Referring to Fig.~\ref{fig:Concept}, we define our system as being comprised of a quantum dot modelled as a two-level system with ground state $|G\rangle$ and excited state $|E\rangle$, a three-level atom with ground state $|g\rangle$ and excited states $|+\rangle$ and $|-\rangle$, and a circulating cavity with two orthogonal modes $a$ and $b$ (counter-clockwise- and clockwise-propagating, respectively), to which both the quantum dot and the atom couple.

To understand the physics behind this scheme, our study begins with a master equation to model the dissipative dynamics of our proposed system. By employing a Born-Markov treatment for the mechanisms of spontaneous emission and cavity decay, we find the following equation of motion for the system density operator $\hat\rho$ ($\hbar=1$):
\begin{align}\label{m_full}
    \begin{split}
        \frac{d\hat{\rho}}{dt} = & -i\commutator{\hat{H}}{\hat{\rho}(t)}+\kappa\left(\mathcal{D}\left[\hat{a}\right] + \mathcal{D}[\hat{b}]\right)\hat{\rho}(t) \\ & \frac{\gamma_q}{2}{D}\left[\hat{\sigma}_{q-}\right]\hat{\rho}(t) + \frac{\gamma_a}{2}\left(\mathcal{D}\left[\hat{\sigma}_{a-}\right] + \mathcal{D}\left[\hat{\sigma}_{b-}\right]\right)\hat{\rho}(t).
    \end{split}
\end{align}
Here, we have the usual Lindblad superoperator defined as $\mathcal{D}[ \hat{O} ]\bullet \equiv 2\hat{O}\bullet\hat{O}^\dagger - \hat{O}^\dagger\hat{O}\bullet - \bullet\hat{O}^\dagger\hat{O}$, the lowering operators $\hat{\sigma}_{q-}\equiv\ket{G}\bra{E}$, $\hat{\sigma}_{a-}\equiv\ket{g}\bra{+}$, and $\hat{\sigma}_{b-}\equiv\ket{g}\bra{-}$, where the former acts on the Hilbert space of the QD and the latter two act on the Hilbert space of the atom. The bosonic annihilation operators $\hat{a}$ and $\hat{b}$ act on the spaces of the degenerate, counterpropagating cavity modes $a$ and $b$, respectively.

Assuming that the quantum dot couples to each cavity mode with strength $g_q$, and that the atom couples to mode $a$ with strength $g_a$ and mode $b$ with strength $g_b$, then in the rotating wave approximation and assuming a Jaynes-Cummings-type interaction between each of the QD and atom with the quantized radiation fields of the cavity, the Hamiltonian may be expressed as
\begin{align}\label{H_full}
    \begin{split}
        \hat{H} = {} & \Delta_c\left( \hat{a}^\dagger\hat{a} + \hat{b}^\dagger\hat{b} \right) \\& + \Delta_q\hat{\sigma}_{q+}\hat{\sigma}_{q-} + \Delta_a\hat{\sigma}_{a+}\hat{\sigma}_{a-} + \Delta_b\hat{\sigma}_{b+}\hat{\sigma}_{b-} \\ & + \left(g_{q}\hat{\sigma}_{q+}\left(\hat{a} + \hat{b}\right) + g_a\hat{\sigma}_{a+}\hat{a} + g_b\hat{\sigma}_{b+}\hat{b} + \text{H.c.}\right) \\ & + \Omega \hat{\sigma}_{q+} + \Omega^*\hat{\sigma}_{q-},
    \end{split}
\end{align}
where H.c. denotes Hermitian conjugate. This Hamiltonian 
is written in a frame rotating at a rate $\omega_L$, corresponding to the frequency of the laser driving the QD with strength $\Omega$. The detunings listed in this equation are therefore defined as the difference between $\omega_L$ and the resonance frequency of the cavity ($\omega_c$), or the transition frequencies between the respective ground and excited states of the two- ($\omega_q$) and three-level emitters ($\omega_\pm$): $\Delta_c\equiv\omega_c-\omega_L$, $\Delta_q\equiv\omega_q-\omega_L$, $\Delta_a\equiv\omega_+-\omega_L$, and $\Delta_b\equiv\omega_--\omega_L$.

\subsection{Cavity Output Fields}

Equations (\ref{m_full}) and (\ref{H_full}) together describe the evolution of a system state in the form of a density operator, $\hat{\rho}(t)$, which is generally mixed. An alternative yet equally useful approach to model the behaviour may be found by working within the Heisenberg picture, where a set of Langevin equations may be derived which describe the temporal evolution of the system operators. The equations of motion for the two cavity annihilation operators are
\begin{subequations}\label{Langevin}
    \begin{align}\label{a(t)}
        \begin{split}
            \frac{d\hat{a}}{dt} & = -i\commutator{\hat{a}(t)}{\hat{H}} - \kappa\hat{a}(t) - \sqrt{2\kappa}\, \hat{a}_\text{in}(t) \\ & = -\left( \kappa + i\Delta_c \right)\hat{a}(t) - ig_q\hat{\sigma}_{q-}(t) - ig_a\hat{\sigma}_{a-}(t) \\ & ~~~~ - \sqrt{2\kappa}\, \hat{a}_\text{in}(t) ,
        \end{split}
    \end{align}
     \begin{align}\label{b(t)}
        \begin{split}
            \frac{d\hat{b}}{dt} & = -i\commutator{\hat{b}(t)}{\hat{H}} - \kappa\hat{b}(t) - \sqrt{2\kappa}\, \hat{b}_\text{in}(t) \\ & = -\left( \kappa + i\Delta_c \right)\hat{b}(t) - ig_q\hat{\sigma}_{q-}(t) - ig_b\hat{\sigma}_{b-}(t) \\ & ~~~~ - \sqrt{2\kappa}\, \hat{b}_\text{in}(t),
        \end{split}
    \end{align}
\end{subequations}
where $\hat{a}_\text{in}(t)$ and $\hat{b}_\text{in}(t)$ are vacuum input field operators. A simple formula may then be obtained which relates the system operators to the fields emitted into and out of the cavity by employing the input-output formalism developed in \cite{Gardiner}:
\begin{subequations}
    \begin{align}
        \hat{a}_{out}(t) = \sqrt{2\kappa}\hat{a}(t) + \hat{a}_{in}(t),
    \end{align}
    \begin{align}
        \hat{b}_{out}(t) = \sqrt{2\kappa}\hat{b}(t) + \hat{b}_{in}(t).
    \end{align}
\end{subequations}
Note, however, that in this work we will only be concerned with properties of the output fields that depend on normally-ordered moments of the output field operators and so the vacuum input field operators will not contribute and can be neglected from here on.

\section{Single excitation regime}
\label{SecSingleEx}

\subsection{Trajectory Analysis}

In this section we wish to analyse the transient dynamics of the system starting from an initial state at time $t=0$ with the quantum dot residing in the excited state, the atom in its ground state and the two cavity modes in the vacuum state. It is assumed that there is no coherent field driving the QD, in which case the Hamiltonian for this system reduces to
\begin{align}\label{H_single}
    \begin{split}
        \hat{H} = {} & \Delta_q\hat{\sigma}_{q+}\hat{\sigma}_{q-} + \Delta_a\hat{\sigma}_{a+}\hat{\sigma}_{a-} + \Delta_b\hat{\sigma}_{b+}\hat{\sigma}_{b-} \\ & + \left(g_{q}\hat{\sigma}_{q+}\left(\hat{a} + \hat{b}\right) + g_a\hat{\sigma}_{a+}\hat{a} + g_b\hat{\sigma}_{b+}\hat{b} + \text{H.c.}\right),
    \end{split}
\end{align}
which is now written in a frame rotating with the cavity resonance frequency, such that $\Delta_q \equiv \omega_q - \omega_c$, $\Delta_a \equiv \omega_+ - \omega_c$ and $\Delta_b \equiv \omega_- - \omega_c$. This Hamiltonian will only couple states included within the one-quantum manifold, while the action of a jump operator ($\hat{a},\hat{b},\hat{\sigma}_{q-},\hat{\sigma}_{a-},\hat{\sigma}_{b-}$) on any of these states will either be zero, or project the system into the (zero excitation) ground state, $\ket{G_0}$. Therefore, one may effectively implement a trajectory unravelling of the master equation, following the methods outlined in \cite{Howard2}, by decomposing the master equation into the sum of two parts,

\begin{align}\label{twoparts}
    \begin{split}
        \frac{d\hat{\rho}}{dt} {} & = \mathcal{L}\hat{\rho}(t) \equiv  \left(\mathcal{M}+\mathcal{N}\right)\hat{\rho}(t),
    \end{split}
\end{align}
with the superoperator $\mathcal{M}$ acting on the one-quantum subspace,

\begin{align}
    \begin{split}
        \mathcal{M} = & -i\commutator{\hat{H}}{\bullet} - \kappa\commutator{\hat{a}^\dagger\hat{a} + \hat{b}^\dagger\hat{b}}{\bullet}_+ - \frac{\gamma_q}{2}\commutator{\hat{\sigma}_{q+}\hat{\sigma}_{q-}}{\bullet}_+ \\ & - \frac{\gamma_a}{2}\commutator{\hat{\sigma}_{a+}\hat{\sigma}_{a-}}{\bullet}_+ - \frac{\gamma_b}{2}\commutator{\hat{\sigma}_{b+}\hat{\sigma}_{b-}}{\bullet}_+,
    \end{split}
\end{align}
where $\commutator{\bullet}{\bullet}_+$ denotes the anticommutator, and the superoperator $\mathcal{N}$  generating transitions to the ground state,
\begin{align}
    \begin{split}
        \mathcal{N} = & 2\kappa(\hat{a}\bullet\hat{a}^\dagger + \hat{b}\bullet\hat{b}^\dagger) + \gamma_q\hat{\sigma}_{q-}\bullet\hat{\sigma}_{q+} \\ & + \gamma_a\hat{\sigma}_{a-}\bullet\hat{\sigma}_{a+} + \gamma_b\hat{\sigma}_{b-}\bullet\hat{\sigma}_{b+}.
    \end{split}
\end{align}
For times $t>0$, the system either emits a photon with a probability of $P(t)$, thus collapsing the state into $\ket{G_0}$, or the system resides in the pure one-quantum state
\begin{align}\label{one_quant}
    \begin{split}
        \ket{\bar{\psi}(t)} = \big{[} & Q(t)\hat{\sigma}_{q+} + A(t)\hat{\sigma}_{a+} \\ & + B(t)\hat{\sigma}_{b+} + \alpha(t)\hat{a}^\dagger + \beta(t)\hat{b}^\dagger\big{]}\ket{G_0},
    \end{split}
\end{align}
where $Q(t)$ is the probability amplitude for the quantum dot to be excited, $A(t)$ ($B(t)$) is the probability amplitude for the state $\ket{+}$ ($\ket{-}$) of the atom to be excited, and $\alpha(t)$ ($\beta(t)$) is the excitation probability amplitude for the $a$-mode ($b$-mode) of the resonator. The density operator is decomposed in a similar manner,
\begin{align}
    \hat{\rho}(t) = P(t)\ket{G_0}\bra{G_0} + (1-P(t))\ket{\bar{\psi}(t)}\bra{\bar{\psi}(t)}.
\end{align}
Here, $\ket{\bar{\psi}(t)}$ evolves according to a non-unitary Schr\"{o}dinger equation,
\begin{align}
    \frac{d\ket{\bar{\psi}}}{dt} = -i\hat{H}_{NH}\ket{\bar{\psi}},
\end{align}
where $\hat H_{NH}$ is the \textit{non-hermitian Hamiltonian}
\begin{align}
    \begin{split}
        \hat{H}_{NH} = & \hat{H} -i\kappa(\hat{a}^\dagger\hat{a} + \hat{b}^\dagger\hat{b}) - i\frac{\gamma_q}{2}\hat{\sigma}_{q+}\hat{\sigma}_{q-} \\ & - i\frac{\gamma_a}{2}\hat{\sigma}_{a+}\hat{\sigma}_{a-} - i\frac{\gamma_b}{2}\hat{\sigma}_{b+}\hat{\sigma}_{b-},
    \end{split}
\end{align}
and the norm of $\ket{\bar{\psi}(t)}$ is equal to $(1-P(t))$. One may then derive the following set of coupled equations for the excited state probability amplitudes:
\begin{subequations}
    \begin{align}
        & \dot{A} = -\left( \gamma_a/2 + i\Delta_a \right)A(t) - ig_a\alpha(t), \label{A}
        \\ & \dot{\alpha} = -\kappa\alpha(t) - ig_qQ(t) - ig_aA(t), \label{alpha}
        \\ & \dot{Q} =  -\left( \gamma_q/2 + i\Delta_q \right)Q(t) - ig_q\alpha(t) - ig_q\beta(t), \label{Q}
        \\ & \dot{\beta} = -\kappa\beta(t) - ig_qQ(t) - ig_bB(t), \label{beta}
        \\ & \dot{B} =  -\left( \gamma_b/2 + i\Delta_b \right)B(t) - ig_b\beta(t). \label{B}
    \end{align}
\end{subequations}

\subsection{Adiabatic Elimination of the Cavity Modes}

As alluded to previously, the focus of this analysis centers on the scheme in which the QD and atom are coupled to the cavity in the bad-cavity, large-cooperativity regime ($\gamma_{q,a}\ll g_{q,a,b}\ll\kappa$). This permits the cavity mode amplitudes in Eqs.~(\ref{A})-(\ref{B}) to be adiabatically eliminated from the dynamics. This is achieved by setting the time derivatives for $\alpha(t)$ and $\beta(t)$ to zero, which yields a set of equations that relate the cavity mode amplitudes to the excited state amplitudes of the QD and atom, i.e.,
\begin{subequations}
    \begin{align}\label{crude2a}
        \alpha(t) = \frac{-i\left(g_qQ(t) + g_aA(t)\right)}{\kappa},
    \end{align}
    \begin{align}\label{crude2b}
        \beta(t) = \frac{-i\left(g_qQ(t) + g_bB(t)\right)}{\kappa}.
    \end{align}
\end{subequations}
Substituting (\ref{crude2a}) and (\ref{crude2b}) into (\ref{A}), (\ref{Q}), and (\ref{B}) gives equations of motion for the QD and atomic excited state probability amplitudes as
\begin{subequations}
    \begin{align}\label{Adot}
        \dot{A} = -\left( \Gamma_a + i\Delta_a \right)A(t) - g_a'Q(t),
    \end{align}
    \begin{align}\label{Qdot}
        \dot{Q} = -\left( \Gamma_q + i\Delta_q \right)Q(t) - g_a'A(t) - g_b'B(t),
    \end{align}
    \begin{align}\label{Bdot}
        \dot{B} = -\left( \Gamma_b + i\Delta_b \right)B(t) - g_b'Q(t),
    \end{align}
\end{subequations}
where we have introduced the cavity-enhanced spontaneous emission rates
\begin{align}
    \begin{split}
        \Gamma_q \equiv \frac{\gamma_q}{2}\left( 1 + \frac{4g_q^2}{\gamma_q\kappa} \right), \quad \Gamma_{a,b} \equiv \frac{\gamma_{a,b}}{2}&\left( 1 + \frac{2 g_{a,b}^2}{\gamma_{a,b}\kappa} \right),
    \end{split}
\end{align}
and the effective coupling strengths between the atom and the quantum dot,
\begin{align}
    g'_{a,b} = \frac{g_qg_{a,b}}{\kappa} .
\end{align}
These couplings between atom and QD are mediated by the cavity modes $a$ and $b$, respectively.

\subsection{Directional Photon Emission: Ideal Case}

We now address an idealized case where relatively simple analytic solutions to Eqs.~(\ref{Adot})-(\ref{Bdot}) may be found, which highlight the mechanism which gives rise to directional photon emission within this system. Here, we assume that $g_b$ - the coupling of the atomic transition to the mode $b$ - is negligible, so that the atom may be modelled as an ideal chiral-two-level system. Additionally, all of the transitions within the QD and atom are assumed to be resonant with the cavity frequency ($\Delta_{q,a,b} = 0$) and the coupling of the quantum dot and atom to the cavity modes far exceeds the spontaneous emission rates, so that we may set $\gamma_{q,a}\approx0$. In this situation Eqs.~(\ref{Adot})-(\ref{Bdot}) reduce to
\begin{subequations}
    \begin{align}\label{Adotideal}
        \dot{A} = -\tilde\Gamma_aA(t) - \sqrt{\frac{\tilde\Gamma_a\tilde\Gamma_q}{2}}Q(t),
    \end{align}
    \begin{align}\label{Qdotideal}
        \dot{Q} = -\tilde\Gamma_qQ(t) - \sqrt{\frac{\tilde\Gamma_a\tilde\Gamma_q}{2}}A(t),
    \end{align}
\end{subequations}
where $\tilde\Gamma_{a} = g_a^2/\kappa$ and $\tilde\Gamma_{q} = 2g_q^2/\kappa$. The solutions to these equations are
\begin{subequations}
    \begin{align}\label{Qapx}
        \begin{split}
            Q(t) = & \frac{1}{2}\left(1 + \frac{\tilde\Gamma_a - \tilde\Gamma_q}{\sqrt{\tilde\Gamma_a^2 + \tilde\Gamma_q^2}}\right)e^{\lambda_+t} \\ & + \frac{1}{2}\left(1 - \frac{\tilde\Gamma_a - \tilde\Gamma_q}{\sqrt{\tilde\Gamma_a^2 + \tilde\Gamma_q^2}}\right)e^{\lambda_-t},
        \end{split}
    \end{align}
    \begin{align}\label{Aapx}
        A(t) = \sqrt{\frac{\tilde\Gamma_q\tilde\Gamma_a}{2\left(\tilde\Gamma_q^2+\tilde\Gamma_a^2\right)}}\left(e^{\lambda_-t} - e^{\lambda_+t}\right),
    \end{align}
\end{subequations}
or, in terms of $g_a$ and $g_q$
\begin{subequations}
    \begin{align}\label{Qapx2}
        \begin{split}
            Q(t) = & \frac{1}{2}\left(1 + \frac{g_a^2 - 2g_q^2}{\sqrt{g_a^4 + 4g_q^4}}\right)e^{\lambda_+t} \\ & + \frac{1}{2}\left(1 - \frac{g_a^2 - 2g_q^2}{\sqrt{g_a^4 + 4g_q^4}}\right)e^{\lambda_-t},
        \end{split}
    \end{align}
    \begin{align}\label{Aapx2}
        A(t) = \frac{g_ag_q}{\sqrt{\left(g_a^4+4g_q^4\right)}}\left(e^{\lambda_-t} - e^{\lambda_+t}\right),
    \end{align}
\end{subequations}
where
\begin{align}
    \lambda_{\pm} = \frac{-(\tilde\Gamma_a+\tilde\Gamma_q)\pm\sqrt{(\tilde\Gamma_a^2+\tilde\Gamma_q^2)}}{2}.
\end{align}
By substituting Eqs.~(\ref{Qapx2}) and (\ref{Aapx2}) into Eqs.~(\ref{crude2a}) and (\ref{crude2b}), the cavity mode amplitudes may be expressed as
\begin{subequations}
    \begin{align}\label{modea}
        \begin{split}
            \alpha(t) = \frac{-ig_q}{2\kappa}\left( \left[ 1 - \frac{g_a^2+2g_q^2}{\sqrt{g_a^4 + 4g_q^4}}\right]e^{\lambda_+t} \right. \\ \left. + \left[ 1 + \frac{g_a^2+2g_q^2}{\sqrt{g_a^4 + 4g_q^4}}\right]e^{\lambda_-t} \right),
        \end{split}
    \end{align}
    \begin{align}\label{modeb}
        \begin{split}
            \beta(t) = \frac{-ig_q}{2\kappa}\left(\left[1 + \frac{g_a^2 - 2g_q^2}{\sqrt{g_a^4 + 4g_q^2}}\right]e^{\lambda_+t} \right. \\ + \left. \left[1 - \frac{g_a^2 - 2g_q^2}{\sqrt{g_a^4 + 4g_q^4}}\right]e^{\lambda_-t}\right).
        \end{split}
    \end{align}
\end{subequations}
The probability of a certain cavity mode emitting a photon may then be calculated using
\begin{subequations}
    \begin{align}\label{Pa}
        P_a = 2\kappa\int_0^\infty dt'\langle(\hat{a}^\dagger\hat{a})(t')\rangle = 2\kappa\int_0^\infty dt'|\alpha(t')|^2,
    \end{align}
    \begin{align}\label{Pb}
        P_b = 2\kappa\int_0^\infty dt'\langle(\hat{b}^\dagger\hat{b})(t')\rangle = 2\kappa\int_0^\infty dt'|\beta(t')|^2.
    \end{align}
\end{subequations}
That is, $P_a$ ($P_b$) is the probability of the system emitting a photon from cavity mode $a$ (mode $b$) at some time during the temporal evolution. We then define the directionality as
\begin{align}\label{Direction1}
    D \equiv \frac{P_b - P_a}{P_b + P_a},
\end{align}
which is a measure of the ability of the system to preferentially emit a photon from the desired cavity mode (in this case, mode $b$).

Using (\ref{modea}) and (\ref{modeb}) the probabilities $P_a$ and $P_b$ are readily evaluated as
\begin{equation}
P_a = \frac{\frac{1}{2}\tilde\Gamma_q}{\tilde\Gamma_a+\tilde\Gamma_q} , ~~~
P_b = \frac{\tilde\Gamma_a+\frac{1}{2}\tilde\Gamma_q}{\tilde\Gamma_a+\tilde\Gamma_q} ,
\end{equation}
which give
\begin{equation}
D = \frac{\tilde\Gamma_a}{\tilde\Gamma_a+\tilde\Gamma_q} .
\end{equation}
These results clearly demonstrate how the presence of a chiral atom can effectively control the emission of the QD into the cavity modes. In particular, if the atomic coupling strength $g_a$ is sufficiently large, such that 
$\tilde\Gamma_a\gg \tilde\Gamma_q$ (i.e., $g_a^2/(2g_q^2)\gg 1$), then $P_b\simeq D \simeq 1$ and the QD emission is predominantly through mode $b$. Note, though, that we still require $\tilde\Gamma_q=2g_q^2/\kappa\gg\gamma_q$ in order for the results of the analysis of this section to hold.

\begin{figure}
    \includegraphics[width=0.5\textwidth]{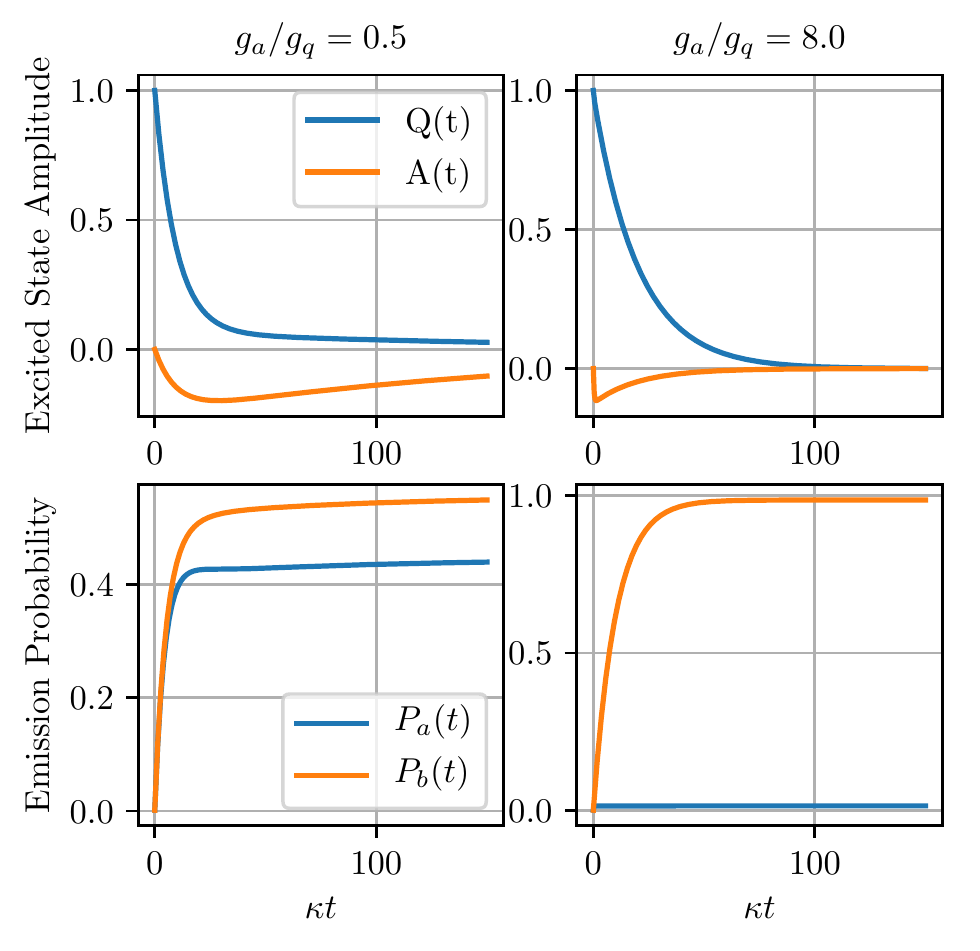}
    \caption{\textbf{Top row:} Excited state amplitudes, Eqs.~(\ref{Qapx2}) and (\ref{Aapx2}), plotted as a function of time for two values of the ratio $g_a/g_q$. \textbf{Bottom row:} Corresponding probability of photon emission from modes $a$ and $b$ as a function of time as given by $P_a(t) = 2\kappa\int_0^t dt'|\alpha(t')|^2$ and $P_b(t) = 2\kappa\int_0^t dt'|\beta(t')|^2$, respectively. Other  parameters are $\{g_q,\Delta_{q,a,b},\gamma_{q,a}\} = \{0.05, 0, 0\}$, expressed in units of $\kappa$}
    \label{Ideal_dynamics}
\end{figure}

Physically, this can be interpreted in terms of destructive quantum interference between QD and atomic dipole fields, i.e., the dipole field associated with the $\ket{g}\leftrightarrow\ket{+}$ atomic transition is $\pi$ out of phase with the $\sigma^+$-polarized component of the QD, and of the same magnitude, leading to a much-diminished amplitude for the $a$-mode field. This is seen explicitly by noting that, in the limit that we consider above, one has $Q(t)\simeq e^{\lambda_+t}$ and $A(t)\simeq -(g_q/g_a)e^{\lambda_+t}$, so $\alpha (t)=-i(g_qQ(t)+g_aA(t))/\kappa\simeq 0$. 
Having a large value of the ratio $g_a/g_q$ means that as soon as some excitation of the QD is transferred to the $a$-mode it is rapidly taken up by the $\ket{g}\leftrightarrow\ket{+}$ atomic transition, enabling destructive interference between the QD and atomic fields over the bulk of the time evolution. 
This behaviour is highlighted in Fig.~\ref{Ideal_dynamics}, where the excited state amplitudes for the QD and the $\ket{g}\leftrightarrow\ket{+}$ atomic transition are plotted as a function of time, along with the photon emission probabilities from the two cavity modes for two values of the ratio $g_a/g_q$  (one small and one large).

\subsection{Third Atomic Level and Spontaneous Emission}

The results obtained above are useful to outline the physical processes within our scheme that cause directional photon emission from the cavity, however the parameter choice is quite removed from what could be achieved within a contemporary experimental setup. In particular, it is unrealistic to neglect the intrinsic process of spontaneous emission from the quantum emitters, as well as the coupling between the $\ket{g}\leftrightarrow\ket{-}$ atomic transition and mode $b$ of the cavity. We therefore shift our attention to a more realistic model, where we now account for these processes and provide a numerical analysis to address their influence on the efficiency of this scheme.

\begin{figure}
    \includegraphics[width=0.5\textwidth]{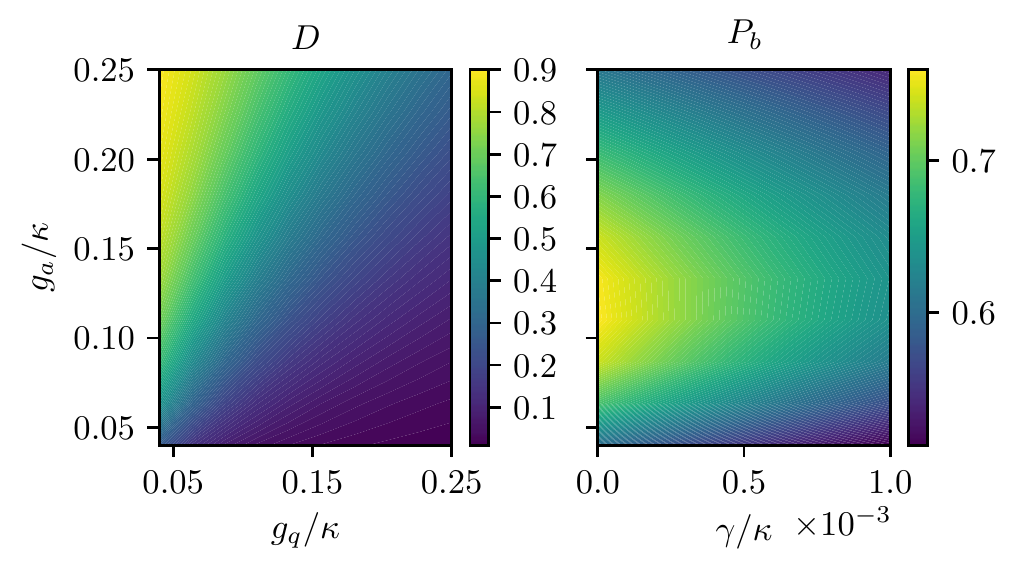}
    \caption{\textbf{Left plot:} Directionality, as defined in Eq.~(\ref{Direction1}), plotted against $g_a$ and $g_q$ for $\gamma/\kappa = 0.001$, $g_b = g_a/\sqrt{45}$ and $\Delta_{q,a,b}/\kappa = 0$. \textbf{Right plot:} Photon emission probability of cavity mode $b$, as defined in Eq.~(\ref{Pb}), plotted against $g_a$ and $\gamma$, with $g_q/\kappa = 0.05$, $g_b = g_a/\sqrt{45}$ and $\Delta_{q,a,b}/\kappa = 0$. Each result shown in this figure has been obtained by solving Eq.~(\ref{m_full}) with the Hamiltonian (\ref{H_single}).}
    \label{3_lev_and_spon}
\end{figure}

To this end, we base our atomic model on a $^{133}\textrm{Cs}$ atom, where the ratio between the coupling strengths of the two transitions to the counter-propagating cavity modes is as large as $g_a/g_b = \sqrt{45}$. For simplicity, the spontaneous emission rate from the excited atomic states is assumed to match that of the QD, i.e., $\gamma_a=\gamma_q \equiv \gamma$. The left-hand plot in Fig.~\ref{3_lev_and_spon} shows a density plot for the directionality, as defined in Eq.~(\ref{Direction1}), against a range of coupling strengths for the QD and atom, when the spontaneous emission rate is set to a value of $\gamma /\kappa= 0.001$. It is observed here that the directionality is optimised when the ratio $g_a/g_q\gg1$, which is consistent with the results obtained for the ideal case where the spontaneous emission rates were neglected. However, the right-hand plot within this figure gives a better representation of how spontaneous emission will influence this scheme. In this plot, the photon emission probability from mode $b$ (Eq.~(\ref{Pb})) is given as a function of the atomic coupling strength and the spontaneous emission rate. It is clear that by increasing the spontaneous emission rates, more of the excitation within the system will be lost to this scattering process. 

Additionally, the emission probability is seen to also decrease above values of $g_a/\kappa\simeq 0.12$. This is due to the fact that upon allowing for a non-zero coupling to the weaker atomic transition, a new eigenstate of the Hamiltonian (\ref{H_single}) arises, which takes the form 
\begin{align}\label{cav_drk}
    \begin{split}
        \ket{CD} \propto \Big( & g_qg_b\ket{G}\ket{+} - g_ag_b\ket{E}\ket{g} \\ & + g_qg_a\ket{G}\ket{-} \Big)\ket{0}_a\ket{0}_b.
    \end{split}
\end{align}
Because the only cavity states that contribute to $\ket{CD}$ are the vacuum states, $\ket{0}_a$ and $\ket{0}_b$, this state will be dark to the cavity modes and will therefore not emit photons into the cavity modes. The overlap between this state and the initial system state $\ket{\psi(t = 0)} = \ket{E}\ket{g}\ket{0}_a\ket{0}_b$ is given by
\begin{align}\label{overlap}
    |\braket{CD}{\psi(t=0)}|^2 = \frac{g_a^2g_b^2}{g_q^2g_b^2 + g_a^2g_b^2 + g_q^2g_a^2}.
\end{align}
This result shows that by increasing the atomic coupling strengths ($g_a$ and $g_b$) the initial state of the system populates more of the cavity-dark eigenstate. As this eigenstate decays only via spontaneous emission, the fraction of emission that is routed into the cavities, and in turn the probability of photon emission into the desired cavity mode, is reduced. A trade-off is therefore identified for the engineering of this system, where an optimal atomic coupling strength should be found which will achieve the highest possible directionality without significantly populating this cavity-dark eigenstate, such that most of the excitation that is initially stored in the QD will be routed into mode $b$ of the cavity. 

However, it is in fact possible for this effect to be reduced by introducing a finite detuning ($\Delta_b$) of the weaker atomic transition from cavity resonance. In this situation, the eigenstate will no longer be dark to the cavity modes, therefore reducing the amount of light that is lost from the system via spontaneous emission. As an example, consider the case corresponding to the top right-hand corner of the plot of $P_b$ in Fig.~\ref{3_lev_and_spon}, where $g_a/\kappa =0.25$, $g_q/\kappa =0.05$, and $\gamma /\kappa =0.001$. With $\Delta_b=0$, we have $P_b=0.54$ and $D=0.87$, but with $\Delta_b/\kappa =0.1$ we obtain $P_b=0.79$ and $D=0.91$, a significant improvement. Furthermore, note that for the parameters of this example $\tilde\Gamma_q/\gamma_q=5$, which is not a lot larger than 1 (so the assumption that $\gamma_q\simeq 0$ is marginal). Doubling the coupling strengths to $g_a/\kappa =0.5$ and $g_q/\kappa =0.1$, and choosing $\Delta_b=0.1$ again, we find further improvement, with $P_b=0.91$ and $D=0.92$. 
We consider the effect of detuning the weakly coupled atomic transition further in the following section, where we examine the regime in which the QD is driven continuously by a weak coherent field and we evaluate the steady-state behaviour of the system.

\section{Weakly driven quantum dot regime}
\label{SecWeaklyDriven}

The analysis of the transient scenario considered in the previous section is useful for highlighting important dynamical aspects of system, although the efficiency of this directional emission scheme is reasonably limited by noise processes that would be typical in practice. In particular, obtaining high values for the directionality, while also maintaining a high emission probability $P_b$ required the consideration of very low spontaneous emission rates of the emitters - much lower than what would be feasible experimentally in the near-future. Here we investigate a slightly altered setup in order to study the behaviour of our system at steady-state, in which the directional emission of the QD into the waveguide is shown to be more robust against the effects of spontaneous emission from the emitters. In contrast to the previous section, we now focus on the situation where the QD is continuously driven by a weak coherent field, which is to say that the parameter $\Omega$ in Eq.~(\ref{H_full}) is non-zero, yet much smaller than the rates $g_q$ and $\kappa$, such that a linearised set of equations may be obtained for the relevant operator expectation values. 

\subsection{Steady-State Output Photon Fluxes}

Starting from Eqs.~(\ref{m_full}) and (\ref{H_full}), the equation of motion for the expectation value of a system operator $\langle\hat{O}\rangle$ follows from the formula $\frac{d\langle\hat{O}\rangle}{dt} = \textrm{Tr}\left[\hat{O}\frac{d\hat{\rho}}{dt}\right]$. The equations of motion for the relevant system operators follow as
\begin{subequations}\label{mean}
    \begin{align}\label{mean1}
        \begin{split}
            \frac{d\langle\hat{\sigma}_{a-}\rangle}{dt} = {} & -\left( i\Delta_a + \frac{\gamma_a}{2} \right)\langle \hat{\sigma}_{a-} \rangle + ig_a\langle \hat{\sigma}_{az} \hat{a} \rangle \\ & + ig_b\langle \hat{\sigma}_{b+}\hat{\sigma}_{a-}\hat{b} \rangle,
        \end{split}
    \end{align}
    \begin{align}\label{mean2}
        \begin{split}
            \frac{d\langle\hat{a}\rangle}{dt} = -\left(i\Delta_c + \kappa\right)\langle \hat{a} \rangle - ig_q\langle \hat{\sigma}_{q-} \rangle - ig_a\langle \hat{\sigma}_{a-} \rangle,
        \end{split}
    \end{align}
    \begin{align}\label{mean3}
        \begin{split}
            \frac{d\langle\hat{\sigma}_{q-}\rangle}{dt} = {} & -\left(i\Delta_q + \frac{\gamma_q}{2}\right)\langle \hat{\sigma}_{q-} \rangle + ig_q\langle \hat{\sigma}_{qz}\hat{a} \rangle \\ & + ig_q\langle \hat{\sigma}_{qz}\hat{b} \rangle + i\Omega\langle \hat{\sigma}_{qz} \rangle,
        \end{split}
    \end{align}
    \begin{align}\label{mean4}
        \begin{split}
            \frac{d\langle\hat{b}\rangle}{dt} = -\left(i\Delta_c + \kappa\right)\langle \hat{b} \rangle - ig_q\langle \hat{\sigma}_{q-} \rangle - ig_b\langle \hat{\sigma}_{b-}\rangle,
        \end{split}
    \end{align}
    \begin{align}\label{mean5}
        \begin{split}
            \frac{d\langle\hat{\sigma}_{b-}\rangle}{dt} = {} & -\left( i\Delta_b + \frac{\gamma_a}{2} \right)\langle \hat{\sigma}_{b-} \rangle + ig_b\langle \hat{\sigma}_{bz} \hat{b} \rangle \\ & + ig_a\langle \hat{\sigma}_{a+}\hat{\sigma}_{b-}\hat{a} \rangle,
        \end{split}
    \end{align}
\end{subequations}
where $\langle\hat{\sigma}_{iz}\rangle = \langle \hat{\sigma}_{i+}\hat{\sigma}_{i-} \rangle - \langle \hat{\sigma}_{i-}\hat{\sigma}_{i+} \rangle$ is the mean inversion of a transition within the QD or atom. These equations are linearised by assuming that the driving strength of the laser, $\Omega$, is sufficiently weak that the quantum dot and atom remain primarily in their ground states. This permits the approximations $\langle\hat{\sigma}_{iz}\rangle \approx -1$, $\langle \hat{\sigma}_{az} \hat{a} \rangle \approx -\langle \hat{a} \rangle$, $\langle \hat{\sigma}_{bz} \hat{b} \rangle \approx -\langle \hat{b} \rangle$ and $\langle \hat{\sigma}_{b+}\hat{\sigma}_{a-}\hat{a} \rangle\approx\langle \hat{\sigma}_{a+}\hat{\sigma}_{b-}\hat{a} \rangle\approx0$. Within this approximation, Eqs.~(\ref{mean1})-(\ref{mean5}) reduce to
\begin{subequations}\label{mean_approx}
    \begin{align}\label{mean_approx1}
        \frac{d\langle\hat{\sigma}_{a-}\rangle}{dt} = -\left( i\Delta_a + \frac{\gamma_a}{2} \right)\langle \hat{\sigma}_{a-} \rangle - ig_a\langle \hat{a} \rangle,
    \end{align}
    \begin{align}\label{mean_approx2}
        \frac{d\langle\hat{a}\rangle}{dt} = -\left(i\Delta_c + \kappa\right)\langle \hat{a} \rangle - ig_q\langle \hat{\sigma}_{q-} \rangle - ig_a\langle \hat{\sigma}_{a-} \rangle,
    \end{align}
    \begin{align}\label{mean_approx3}
        \frac{d\langle\hat{\sigma}_{q-}\rangle}{dt} = -\left(i\Delta_q + \frac{\gamma_q}{2}\right)\langle \hat{\sigma}_{q-} \rangle - ig_q\langle \hat{a} \rangle - ig_q\langle \hat{b} \rangle - i\Omega,
    \end{align}
    \begin{align}\label{mean_approx4}
        \frac{d\langle\hat{b}\rangle}{dt} = -\left(i\Delta_c + \kappa\right)\langle \hat{b} \rangle - ig_q\langle \hat{\sigma}_{q-} \rangle - ig_b\langle \hat{\sigma}_{b-}\rangle,
    \end{align}
    \begin{align}\label{mean_approx5}
        \frac{d\langle\hat{\sigma}_{b-}\rangle}{dt} = -\left( i\Delta_b + \frac{\gamma_a}{2} \right)\langle \hat{\sigma}_{b-} \rangle - ig_b\langle \hat{b} \rangle.
    \end{align}
\end{subequations}
We are interested in the steady-state behaviour. Therefore we set the time derivatives in Eqs.~(\ref{mean_approx1})-(\ref{mean_approx5}) to zero. For $\gamma = \gamma_q = \gamma_a$ and $\Delta_{c,q,a,b} = 0$, a relatively simple set of algebraic equations are found for the following expectation values of the system operators at steady-state:
\begin{subequations}\label{ss}
    \begin{align}\label{ssq}
        \langle \hat{\sigma}_{q-} \rangle_{ss} = \frac{-2i\Omega}{\gamma\left(1 + 2C_q\left[\frac{1}{1+2C_a} + \frac{1}{1+2C_b}\right]\right)},
    \end{align}
    \begin{align}\label{ssA}
        \langle \hat{\sigma}_{a-} \rangle_{ss} = \frac{-2g_qg_a}{\kappa\gamma(1+2C_a)}\langle \hat{\sigma}_{q-} \rangle_{ss},
    \end{align}
    \begin{align}\label{ssB}
        \langle \hat{\sigma}_{b-} \rangle_{ss} = \frac{-2g_qg_b}{\kappa\gamma(1+2C_b)}\langle \hat{\sigma}_{q-} \rangle_{ss},
    \end{align}
    \begin{align}\label{ssa}
        \langle \hat{a} \rangle_{ss} = \frac{-ig_q}{\kappa(1+2C_a)}\langle \hat{\sigma}_{q-} \rangle_{ss},
    \end{align}
    \begin{align}\label{ssb}
        \langle \hat{b} \rangle_{ss} = \frac{-ig_q}{\kappa(1+2C_b)}\langle \hat{\sigma}_{q-} \rangle_{ss}.
    \end{align}
\end{subequations}
Here we have defined the cooperativities $C_q\equiv g_q^2/(\kappa\gamma)$, $C_a\equiv g_a^2/(\kappa\gamma)$ and $C_b\equiv g_b^2/(\kappa\gamma)$. Eqs.~(\ref{ssq})-(\ref{ssb}) reveal how directional emission is induced in the system at steady-state. It is apparent that both $\langle \hat{\sigma}_{a-} \rangle_{ss}$ and $\langle \hat{\sigma}_{b-} \rangle_{ss}$ are perfectly out of phase with $\langle \hat{\sigma}_{q-} \rangle_{ss}$, which again implies that the radiation from the quantum dot will interfere destructively with the re-radiation from the atom; this is essentially the same effect that was found to give rise to the directional photon emission that was addressed in Section \ref{SecSingleEx}. As the atomic coupling to mode $a$ of the cavity is much larger than that for mode $b$, this interference effect will be larger within mode $a$ and emission from this mode will be largely suppressed. 



\begin{figure}{h}
    \includegraphics[width=1.1\columnwidth]{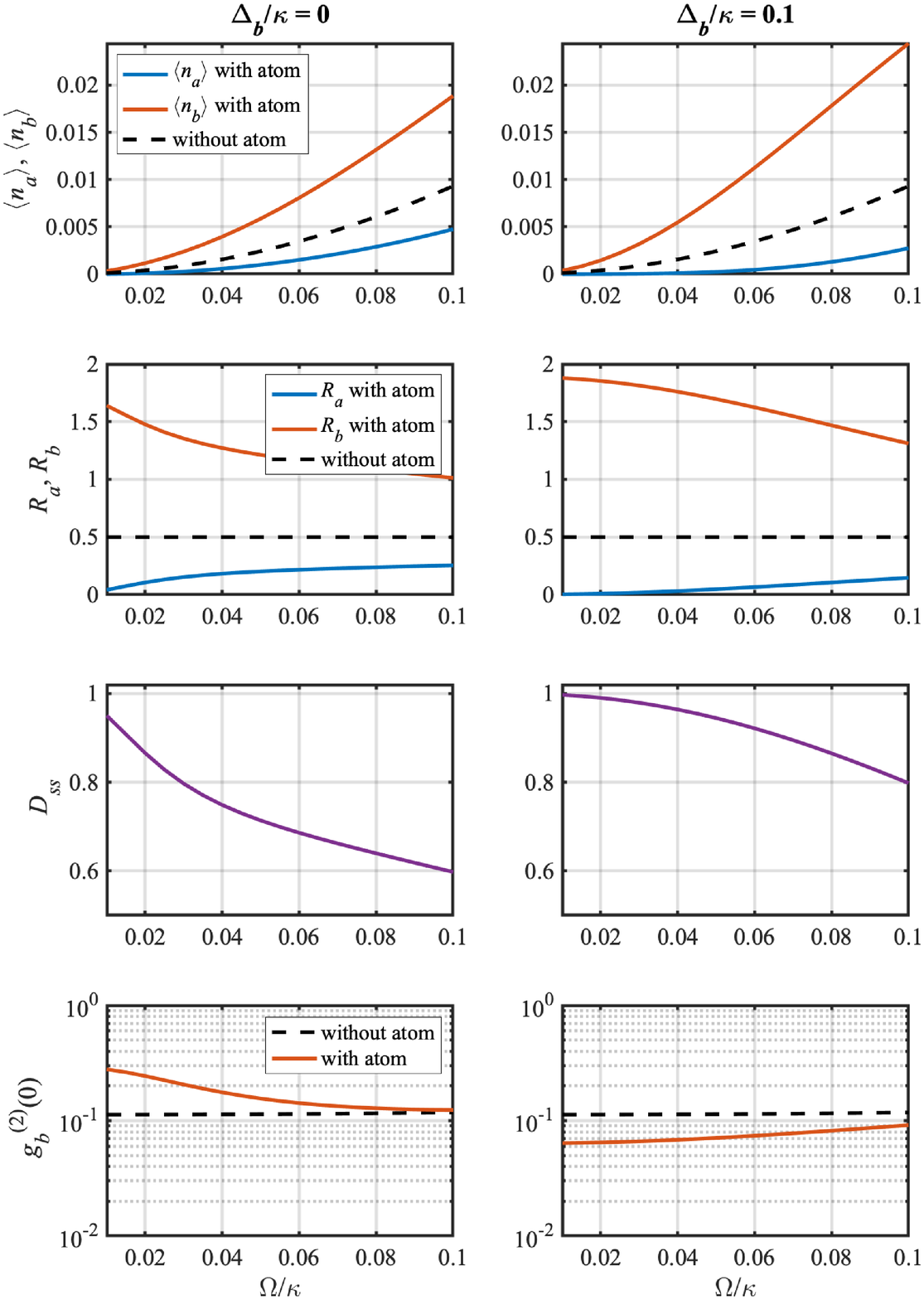}   \\
    \vspace{-8mm}
    \caption{Mean steady-state intracavity photon numbers $\langle n_a\rangle$ and $\langle n_b\rangle$, ratios $R_a$ and $R_b$, directionality $D_{ss}$, and second-order intensity correlation function $g_b^{(2)}(0)$ as a function of the driving strength of the QD for $\Delta_b/\kappa=0$ (left column) and $\Delta_b/\kappa=0.1$ (right column). Other parameters are $\{g_q,g_a,g_b,\gamma_q,\gamma_a\}/\kappa =\{ 0.5,0.5,0.5/\sqrt{45},0.01,0.01\}$ and $\Delta_{c,q,a}=0$. All results shown in this figure have been obtained by solving Eq.~(\ref{m_full}) with the Hamiltonian (\ref{H_single}).
    }
    \label{varying_driving1}
\end{figure}

To study this effect and others in more detail analytically, we can look more carefully at the results (\ref{ssq})-(\ref{ssb}) in certain limits of interest. First, let us consider the case of no atom ($C_a=C_b=0$) and $2C_q\gg 1$. Then
\begin{align}
\langle \hat{\sigma}_{q-} \rangle_{ss}^0 = -\frac{2i\Omega}{\gamma (1+4C_q)} \simeq -\frac{i\Omega}{2\gamma C_q} ,
\end{align}
and
\begin{align}
\langle \hat{a} \rangle_{ss}^0 = \langle \hat{b} \rangle_{ss}^0 = - \frac{ig_q}{\kappa} \langle \hat{\sigma}_{q-} \rangle_{ss}^0 \simeq - \frac{\Omega}{2g_q} .
\end{align}
The output photon fluxes from the two cavity modes are then 
\begin{align}
\Phi_{a,ss}^0 &= 2\kappa\langle \hat a^\dag \hat a\rangle_{ss}^0 \simeq 2\kappa |\langle \hat{a} \rangle_{ss}^0|^2 \simeq \frac{\Omega^2}{4g_q^2} , 
\\
\Phi_{b,ss}^0 &= \Phi_{a,ss}^0 \simeq \frac{\Omega^2}{4g_q^2} .
\end{align}
Now, consider the case in which the atom is coupled with $2C_a\gg 1$ and $C_q,C_a\gg C_b$ (and, again, $2C_q\gg 1$). We find
\begin{align}
\langle \hat{\sigma}_{q-} \rangle_{ss} \simeq -\frac{i\Omega}{2\gamma C_q} 2(1+2C_b) .
\end{align}
We note immediately that the QD polarization is a factor of $2(1+2C_b)$ {\em larger} than for the no-atom case. Meanwhile, for the cavity field amplitudes we find
\begin{align}
\langle \hat{a} \rangle_{ss} \simeq - \frac{\Omega}{g_q} \frac{1+2C_b}{2C_a} \simeq 0 ,
\end{align}
and
\begin{align}
\langle \hat{b} \rangle_{ss} \simeq - \frac{\Omega}{g_q} .
\end{align}
The amplitude $\langle \hat{b} \rangle_{ss}$ is a factor of 2 larger than for the no-atom case. This means that the (coherent) output photon flux from mode $b$ is enhanced by a factor of 4 over its corresponding no-atom value and, further, that the total output photon flux from the system through the cavity modes is, for a given driving strength of the QD, {\em doubled}, i.e., $\Phi_{a,ss}+\Phi_{b,ss}=2(\Phi_{a,ss}^0+\Phi_{b,ss}^0)$. Moreover, the directionality, defined for the continuous driving case by
\begin{align}\label{Direction2}
    D_{ss} \equiv \frac{\Phi_{b,ss} - \Phi_{a,ss}}{\Phi_{b,ss} + \Phi_{a,ss}},
\end{align}
is essentially equal to 1 in the limit considered.

To back up these approximate analytical calculations, in Fig.~\ref{varying_driving1} we plot the steady-state intracavity photon numbers $\langle n_a\rangle=\langle\hat a^\dag\hat a\rangle$ and $\langle n_b\rangle=\langle\hat b^\dag\hat b\rangle$ (equivalent to the output photon fluxes from the two modes scaled by $2\kappa$), and the directionality $D_{ss}$, computed numerically from the full master equation (\ref{m_full}), as a function of the driving strength $\Omega$ of the QD and for two values of the detuning of the atomic transition $|g\rangle\leftrightarrow |-\rangle$ from the driving laser frequency, $\Delta_b=\omega_--\omega_L$. We also plot the ratios of the output fluxes from each cavity mode in the presence of the atom to the total output flux from both modes in the absence of the atom, i.e., we plot
\begin{align}\label{R_ab}
    R_a = \frac{\Phi_{a,ss}}{\Phi_{a,ss}^0 + \Phi_{b,ss}^0} , ~~~ R_b = \frac{\Phi_{b,ss}}{\Phi_{a,ss}^0 + \Phi_{b,ss}^0} .
\end{align}
Finally, we plot the 2nd-order intensity correlation function $g_b^{(2)}(0)$, which we will discuss in more detail in the next section.

For the parameters of Fig.~\ref{varying_driving1}, we have $C_q=C_a=25$ and $C_b=0.556$. We see that there is good qualitative agreement with the predictions of the approximate analysis given above. In particular, the output photon flux from mode $b$ ($a$) is substantially enhanced (reduced) with the addition of the atom. Furthermore, this enhancement does indeed take it well beyond the no-atom total output flux over much of the range of driving strengths considered, while good directionality is also demonstrated. 

These effects are noticeably further enhanced with the addition of a finite detuning, $\Delta_b/\kappa =0.1$, leading also to better quantitative agreement with the predictions of the linear analysis above. We note that for $\Delta_b/\kappa =0$ the numerical solutions of the master equation reveal a significant population of the atomic state $|-\rangle$ for the driving strengths considered. With $\Delta_b/\kappa =0.1$, however, this population becomes negligible. Some insight is offered by modifying the analysis above to allow for finite $\Delta_b$. In particular, doing so one finds, for $2\Delta_b/\gamma_a\gg 1$ (in Fig.~\ref{varying_driving1}, for $\Delta_b/\kappa =0.1$ we have $2\Delta_b/\gamma_a=20$), that $\langle\hat b\rangle_{\rm ss}\simeq -\Omega /g_q$ once again, but
\begin{align}
\langle\hat\sigma_{b-}\rangle_{ss} = -\frac{ig_b}{i\Delta_b+\gamma_a/2} \langle\hat b\rangle_{\rm ss} \simeq \frac{g_b\Omega}{g_q\Delta_b} \ll \langle\hat\sigma_{b-}\rangle_{ss}^{\Delta_b=0} .
\end{align}

\begin{figure}{h}
    \includegraphics[width=1.1\columnwidth]{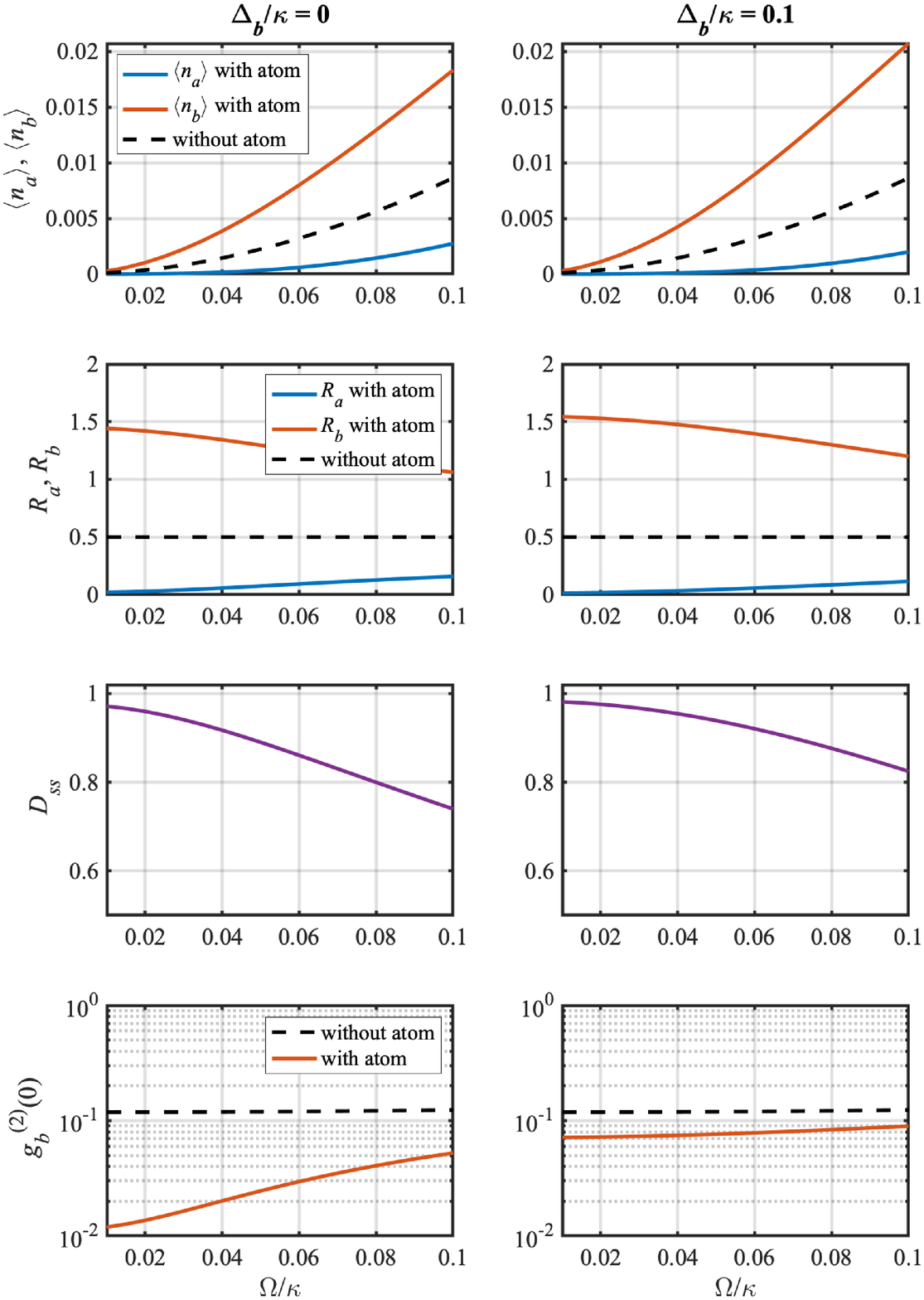}   \\
    \vspace{-8mm}
    \caption{Mean steady-state intracavity photon numbers $\langle n_a\rangle$ and $\langle n_b\rangle$, ratios $R_a$ and $R_b$, directionality $D_{ss}$, and second-order intensity correlation function $g_b^{(2)}(0)$ as a function of the driving strength of the QD for $\Delta_b/\kappa=0$ (left column) and $\Delta_b/\kappa=0.1$ (right column). Other parameters are the same as for Fig.~\ref{varying_driving1}, except that now $\{\gamma_q,\gamma_a\}/\kappa =\{ 0.05,0.05\}$. All results shown in this figure have been obtained by solving Eq.~(\ref{m_full}) with the Hamiltonian (\ref{H_single}).
    }
    \label{varying_driving2}
\end{figure}

The effects described above for weak, continuous driving -- enhanced photon flux from mode $b$ and good directionality -- are also more robust with respect to QD and atomic spontaneous emission than the single-photon pulse regime of the previous section. This is demonstrated in Fig.~\ref{varying_driving2}, where the results shown are obtained for the same parameters as in Fig.~\ref{varying_driving1} except that now we set $\gamma_q/\kappa=\gamma_a/\kappa =0.05$. Note that for these values we have $C_q=C_a=5$. (Note also that fairly similar results are obtained with still larger spontaneous emission rates, i.e., with $\gamma_q/\kappa=\gamma_a/\kappa =0.1$, for which $C_q=C_a=2.5$.)

\subsection{Photon Statistics and Correlations}
\label{SecPhotonStats}

\subsubsection{Weak Driving Approximation}

Our analysis is concluded by briefly addressing the photon statistics of the fields emitted from the cavity. We are therefore interested in evaluating the following autocorrelation functions
\begin{subequations}
    \begin{align}\label{2t1}
        g_a^{(2)}(t,\tau) = \frac{\langle \hat{a}^\dagger(t)\hat{a}^\dagger(t+\tau)\hat{a}(t+\tau)\hat{a}(t)\rangle}{\langle \hat{a}^\dagger (t)\hat{a}(t)\rangle^2},
    \end{align}
    \begin{align}\label{2t2}
        g_b^{(2)}(t,\tau) = \frac{\langle \hat{b}^\dagger(t)\hat{b}^\dagger(t+\tau)\hat{b}(t+\tau)\hat{b}(t)\rangle}{\langle \hat{b}^\dagger (t)\hat{b}(t)\rangle^2},
    \end{align}
\end{subequations}
in the steady-state limit $t\xrightarrow{}\infty$. Eq.~(\ref{2t1}) ((\ref{2t2})) then gives the relative change in likelihood for the detection of a photon  from mode $a$ ($b$) of the cavity at time $\tau$ later than an initial photon detection from mode $a$ ($b$) at time $t$. To simplify notation, the parameter $t$ is dropped from Eqs.~(\ref{2t1}) and (\ref{2t2}) where it should be understood that the system resides in its steady state at time $\tau=0$. We again specialise to the bad-cavity regime and assume the driving laser is on resonance with the cavity modes, the quantum dot and both atomic transitions ($\Delta_{c,q,a,b} = 0$). Additionally, the assumption is made that the driving laser is weak enough that the steady-state of the system is approximately a pure state. This allows for a rather straightforward evaluation of Eqs.~(\ref{2t1}) and (\ref{2t2}) at zero time delay ($\tau=0$), which follows from the methods outlined in Section (13.2.3) of \cite{Howard2}.

Working within these parameter constraints, it is possible to again adiabatically eliminate the cavity modes from the system dynamics. At the level of the master equation, this may be achieved by tracing over the cavity modes in Eq.~(\ref{m_full}) to obtain a master equation for the reduced density operator $\hat{\rho}_r(t)$ that describes an effective interaction between the QD and atom:
\begin{align}
    \begin{split}
        \frac{d\hat{\rho}_r}{dt} = & -i\Omega\commutator{\hat{\sigma}_{q+} + \hat{\sigma}_{q-}}{\hat{\rho}_r} + \frac{\gamma_q}{2}\mathcal{D}\left[\hat{\sigma}_{q-}\right]\hat{\rho}_r \\ & + \frac{\gamma_a}{2}\left(\mathcal{D}\left[\hat{\sigma}_{a-}\right] + \mathcal{D}\left[\hat{\sigma}_{b-}\right]\right)\hat{\rho}_r \\ & + \frac{1}{\kappa}\left(\mathcal{D}\left[g_q\hat{\sigma}_{q-} + g_a\hat{\sigma}_{a-}\right] + \mathcal{D}\left[g_q\hat{\sigma}_{q-} + g_b\hat{\sigma}_{b-}\right]\right)\hat{\rho}_r .
    \end{split}
\end{align}
The adiabatically eliminated cavity operators (in the Heisenberg picture) are expressed in terms of operators acting within the Hilbert spaces of the QD and atom 
\begin{subequations}
    \begin{align}
        & \hat{a}(t) = \frac{-i}{\kappa}\left( g_q\hat{\sigma}_{q-}(t) + g_a\hat{\sigma}_{a-}(t) \right) + \text{v.f.}, \label{a_elim}\\ &
        \hat{b}(t) = \frac{-i}{\kappa}\left( g_q\hat{\sigma}_{q-}(t) + g_b\hat{\sigma}_{b-}(t) \right) + \text{v.f.}, \label{b_elim}
    \end{align}
\end{subequations}
where v.f. denotes the vacuum field contribution, which has again been included for completeness, but may be neglected for this analysis. With a sufficiently weak driving laser strength $\Omega$, the state of the system can be expanded as a pure state,
\begin{align}\label{two_quant}
    \begin{split}
        \ket{\psi(t)} = & \ket{G,g} + \alpha(t)\ket{E,g} + \beta(t)\ket{G,+} \\ & + \eta(t)\ket{G,-} + \zeta(t)\ket{E,+} + \xi(t)\ket{E,-},
    \end{split}
\end{align}
where we allow for up to two quanta of excitation. Note that the amplitudes $\alpha(t)$ and $\beta(t)$ are not the same as those shown in Eq.~(\ref{one_quant}). The state (\ref{two_quant}) will evolve according to a non-unitary Schr\"{o}dinger equation,
\begin{align}\label{non_unit}
    \frac{d\ket{\psi}}{dt} = -i\hat{\mathcal{H}}\ket{\psi},
\end{align}
with the non-Hermitian Hamiltonian
\begin{align}
    \begin{split}
        \hat{\mathcal{H}} = & \Omega\hat{\sigma}_{q+} - \frac{i\gamma_q}{2}\hat{\sigma}_{q+}\hat{\sigma}_{q-} - \frac{i\gamma_a}{2}\hat{\sigma}_{a+}\hat{\sigma}_{a-} - \frac{i\gamma_a}{2}\hat{\sigma}_{b+}\hat{\sigma}_{b-} \\ & - \frac{i}{\kappa}\left( g_q\hat{\sigma}_{q+} + g_a\hat{\sigma}_{a+} \right)\left( g_q\hat{\sigma}_{q-} + g_a\hat{\sigma}_{a-} \right) \\ & - \frac{i}{\kappa}\left( g_q\hat{\sigma}_{q+} + g_b\hat{\sigma}_{b+} \right)\left( g_q\hat{\sigma}_{q-} + g_b\hat{\sigma}_{b-} \right).
    \end{split}
\end{align}
From this, we may derive equations of motion for the complex amplitudes appearing in Eq.~(\ref{two_quant}), which take the forms

\begin{subequations}
\begin{align}
\dot\alpha &= -\Gamma_q\alpha - \frac{g_q}{\kappa} (g_a\beta + g_b\eta ) ,
\\
\dot\beta &= -\Gamma_a\beta - \frac{g_ag_q}{\kappa}\alpha ,
\\
\dot\eta &= -\Gamma_b\eta - \frac{g_bg_q}{\kappa}\alpha ,
\\
\dot\zeta &= -(\Gamma_q+\Gamma_a)\zeta - i\Omega\beta ,
\\
\dot\xi &= -(\Gamma_q+\Gamma_b)\xi - i\Omega\eta ,
\end{align}
\end{subequations}
where $\Gamma_q=\frac{\gamma_q}{2}(1+4C_q)$, $\Gamma_a=\frac{\gamma_a}{2}(1+2C_a)$, and $\Gamma_b=\frac{\gamma_a}{2}(1+2C_b)$. Solving these equations at steady-state yields
\begin{subequations}
    \begin{align}
        & \alpha_{ss} = \frac{-i\Omega}{\Gamma_q - \frac{g_q^2}{\kappa^2}\left(\frac{g_a^2}{\Gamma_a} + \frac{g_b^2}{\Gamma_b}\right)},
        \\ & \beta_{ss} = -\frac{g_ag_q}{\Gamma_a\kappa}\alpha_{ss},
        \\ & \eta_{ss} = -\frac{g_bg_q}{\Gamma_b\kappa}\alpha_{ss},
        \\ & \zeta_{ss} = \frac{i\Omega}{\left( \Gamma_q + \Gamma_a \right)}\frac{g_ag_q}{\Gamma_a\kappa}\alpha_{ss},
        \\ & \xi_{ss} = \frac{i\Omega}{\left( \Gamma_q + \Gamma_b\right)}\frac{g_bg_q}{\Gamma_b\kappa}\alpha_{ss} .
    \end{align}
\end{subequations}

Within this approximation, the autocorrelation functions (\ref{2t1}) and (\ref{2t2}) are given by
\begin{subequations}
    \begin{align}
        & g_a^{(2)}(\tau) = \frac{\bra{\psi_a(\tau)}\hat a^\dagger \hat a\ket{\psi_a(\tau)}}{\bra{\psi_{ss}}\hat a^\dagger \hat a\ket{\psi_{ss}}}, \\ & g_b^{(2)}(\tau) = \frac{\bra{\psi_b(\tau)}\hat b^\dagger \hat b\ket{\psi_b(\tau)}}{\bra{\psi_{ss}}\hat b^\dagger \hat b\ket{\psi_{ss}}},
    \end{align}
\end{subequations}
where $\ket{\psi_{ss}}$ is the state vector (\ref{two_quant}) at steady-state and $\ket{\psi_{a,b}(\tau)}$ are found by solving Eq.~(\ref{non_unit}) subject to the initial condition
\begin{subequations}\label{ab}
    \begin{align}
        \label{init1} & \ket{\psi_a(\tau = 0)} = \frac{\hat a\ket{\psi_{ss}}}{\sqrt{\bra{\psi_{ss}}\hat a^\dagger \hat a\ket{\psi_{ss}}}}, \\ &
        \label{init2} \ket{\psi_b(\tau = 0)} = \frac{\hat b\ket{\psi_{ss}}}{\sqrt{\bra{\psi_{ss}}\hat b^\dagger \hat b\ket{\psi_{ss}}}},
    \end{align}
\end{subequations}
i.e., the steady-state, $\ket{\psi_{ss}}$, conditioned on the emission of a photon from mode $a$ and mode $b$ of the cavity, respectively. Note that in (\ref{ab}), we take $\hat a = -i(g_q\hat\sigma_{q-}+g_a\hat\sigma_{a-})/\kappa$ and $\hat b = -i(g_q\hat\sigma_{q-}+g_b\hat\sigma_{b-})/\kappa$.

\subsubsection{Zero Time Delay ($\tau = 0$)}

Explicit expressions for the correlation functions are now derived in the limit $\tau=0$. Writing out Eqs.~(\ref{init1}) and (\ref{init2}) explicitly, we find
\begin{subequations}
    \begin{align}\label{conditional1}
        \begin{split}
            \ket{\psi_a(\tau = 0)} = \frac{1}{\sqrt{n_{a}}}\bigg[ & \left( \frac{g_q}{\kappa}\alpha_{ss} + \frac{g_a}{\kappa}\beta_{ss} \right)\ket{G,g} \\ & + \frac{g_q}{\kappa}\zeta_{ss}\ket{G,+} + \frac{g_q}{\kappa}\xi_{ss}\ket{G,-} \\ & + \frac{g_a}{\kappa}\zeta_{ss}\ket{E,g} \bigg],
        \end{split}
    \end{align}
    \begin{align}\label{conditional2}
        \begin{split}
            \ket{\psi_b(\tau = 0)} = \frac{1}{\sqrt{n_{b}}}\bigg[ & \left( \frac{g_q}{\kappa}\alpha_{ss} + \frac{g_b}{\kappa}\eta_{ss} \right)\ket{G,g} \\ & + \frac{g_q}{\kappa}\zeta_{ss}\ket{G,+} + \frac{g_q}{\kappa}\xi_{ss}\ket{G,-} \\ & + \frac{g_b}{\kappa}\xi_{ss}\ket{E,g} \bigg],
        \end{split}
    \end{align}
\end{subequations}
with
\begin{subequations}
    \begin{align}
        \begin{split}
            n_{a} \equiv & \frac{g_q^2}{\kappa^2}|\alpha_{ss} +(g_a/g_q)\beta_{ss}|^2 \\ & + \left( \frac{g_q^2 + g_a^2}{\kappa^2} \right)|\zeta_{ss}|^2 + \frac{g_q^2}{\kappa^2}|\xi_{ss}|^2,
        \end{split}
    \end{align}
and
    \begin{align}
        \begin{split}
            n_{b} \equiv & \frac{g_q^2}{\kappa^2}|\alpha_{ss} + (g_b/g_q)\eta_{ss}|^2 \\ & + \left( \frac{g_q^2 + g_b^2}{\kappa^2} \right)|\xi_{ss}|^2 + \frac{g_q^2}{\kappa^2}|\zeta_{ss}|^2.
        \end{split}
    \end{align}
\end{subequations}
From this, we obtain the \textit{autocorrelation functions with zero time delay},
\begin{subequations}
    \begin{align}
        & g_a^{(2)}(0) = \frac{g_a^2g_q^2}{\kappa^4}\frac{|\zeta_{ss}|^2}{n_{a}^2}, \label{0time1}
        \\ & g_b^{(2)}(0) = \frac{g_b^2g_q^2}{\kappa^4}\frac{|\xi_{ss}|^2}{n_{b}^2}, \label{0time2}
    \end{align}
\end{subequations}
which, using the results above in the limit $\Omega\rightarrow 0$, reduce to
\begin{widetext}
\begin{subequations}
    \begin{align}
        g_a^{(2)}(0) &= C_a^2(1+2C_a)^2 \frac{\gamma_q^2}{(\Gamma_q+\Gamma_a)^2} \left[ 1 + 2C_q \left( \frac{1}{1+2C_a} + \frac{1}{1+2C_b} \right) \right]^2 ,
        \label{0time1b}
        \\ 
        g_b^{(2)}(0) &= C_b^2(1+2C_b)^2 \frac{\gamma_q^2}{(\Gamma_q+\Gamma_b)^2} \left[ 1 + 2C_q \left( \frac{1}{1+2C_a} + \frac{1}{1+2C_b} \right) \right]^2 .
        \label{0time2b}
    \end{align}
\end{subequations}
\end{widetext}
Within the regime of most interest to us here, i.e., $2C_q\simeq 2C_a\gg 1>C_b$, these results predict extreme bunching of photons emitted from mode $a$ ($g_a^{(2)}(0)\gg 1$) and antibunching of photons emitted from mode $b$ ($g_b^{(2)}(0)< 1$). This extreme difference in the photon statistics of the two modes can be explained by examining Eqs.~(\ref{a_elim}) and (\ref{b_elim}). In particular, upon the event of a photon emission, it is uncertain whether the photon was emitted from the quantum dot or from the atom, meaning that the system resides in an entangled state, i.e., a coherent superposition of the ground state $\ket{G,g}$ and the three excited states $\ket{E,g}$, $\ket{G,+}$ and $\ket{G,-}$. 
Due to the much larger coupling of the atom to mode $a$, one has $|g_a\zeta_{ss}|\gg|g_b\xi_{ss}|$ and the population of the excited states is larger within the conditional state $\ket{\psi_a(\tau = 0)}$ than for $\ket{\psi_b(\tau = 0)}$. Furthermore, for the amplitude of the ground state $|G,g\rangle$ in $\ket{\psi_a(\tau = 0)}$ one finds
\begin{align}
\frac{g_q}{\kappa}\alpha_{ss} + \frac{g_a}{\kappa}\beta_{ss} = \frac{g_q}{\kappa} \frac{1}{1+2C_a} \alpha_{ss} \simeq 0
\end{align}
for $2C_a\gg 1$ and so the excited state amplitudes dominate in $\ket{\psi_a(\tau = 0)}$, leading to strong bunching in mode $a$. In contrast, the amplitude of the ground state $|G,g\rangle$ in $\ket{\psi_b(\tau = 0)}$ is in fact the dominant amplitude in the state, thus giving rise to antibunching in mode $b$.

The numerical solutions of the full master equation confirm these bunching and antibunching behaviours of the $a$ and $b$ modes, respectively, in the appropriate parameter regime ($\kappa >g_{q,a,b}$, $C_{q,a}\gg C_b$). Strong antibunching of mode $b$ is illustrated by the plots of $g_b^{(2)}(0)$ versus $\Omega /\kappa$ in Figs.~\ref{varying_driving1} and \ref{varying_driving2}. The corresponding $g_a^{(2)}(0)$ is not plotted as it takes values several orders of magnitude larger than $g_b^{(2)}(0)$.



\section{Discussion}
\label{SecDisc}
We have demonstrated that directional coupling from an unpolarized emitter to a circulating cavity can be induced by coupling another emitter chirally to the same cavity. Our analysis reveals that in both the single excitation regime and the steady state of the driven regime, the mechanism behind the directionality is a quantum interference effect, wherein the field amplitude of one of the modes (and therefore in one direction) is strongly suppressed. For an ideal system, and assuming the coupling asymmetry of atomic cesium,  directionalities well in excess of 90$\%$ can be achieved. Furthermore, in the continuous driving case we find that the intensity of the directional emission is also significantly enhanced compared to emission from the QD in the absence of the atom, and that the directional emission is still strongly antibunched.

We now make several comments related to the scheme presented above.
Firstly, 
an important point to consider is the experimental realizability of the proposal. In recent years, a number of experimental studies have demonstrated large Purcell factors, typically for the case of photonic crystal cavities~\cite{Englund}. Although resonators with circulating geometries have not typically been used for such demonstrations, there is no a priori reason why large Purcell enhancement cannot be achieved in such a configuration, and the large quality factors and relatively small mode volumes necessary have already been demonstrated in a number of cases~\cite{AokiResonator, RauschResonator}. In addition, because the scheme presented here works in the bad cavity regime, it should also be possible to use it with plasmonic resonators which typically exhibit very fast decay times (large $\kappa$) along with large coupling rates.

Another important point to note, as mentioned in the introduction, is that there is no need for the scheme to use just a single atom. Indeed, collective coupling of optically pumped atoms to the resonator can alleviate the need for large single atom couplings, due to the collective enhancement factor of $\sqrt{N}$ which is applied to the single atom coupling rate~\cite{ScottMaarten}, and at the level of a single-excitation, or for weak continuous driving, the response of an atomic ensemble is the same as for a single atom. Additionally, the application of a magnetic field as shown in Fig.~\ref{fig:Concept} is not strictly necessary to stabilise the atomic spin states, as it was recently shown in \cite{pucher2021atomic} that chiral atomic coupling can also be implemented using tensor light shifts.

More speculatively, it may be possible to further simplify the setup by using emitters which have structural related chirality at 
room temperature. Particles such as carbon nanotubes~\cite{Wei,Sato} and transition metal dichalcogenides~\cite{Hu} exhibit circular dipole moments at room temperature, but are not in general single photon emitters. It might be possible, therefore, to replace the atoms in our current work with such nanomaterials. Indeed, research regarding the coupling of these materials is already underway~\cite{Hu,Khas,MarkCNT}.

In summary, directional emission of single photons enabled by chiral coupling between quantum emitters and nanophotonic devices is a technique with important applications to future quantum information technologies. However, given that it requires a polarized emitter dipole moment, its use is typically restricted to ultra-cold systems. Nonetheless, as we show here, by using a circulating cavity, directional emission can effectively be transferred from a chirally coupled emitter to a randomly polarized emitter, even when both emitters are coupled to the resonator in the bad-cavity regime. We anticipate that this result may allow the easing of requirements on the types of emitters which can be used in directional emission schemes.

\subsection*{Acknowledgments}
MS acknowledges funding from JPS Kakenhi , a Matsuo Foundation grant, and funding from the Quantum Nanophotonic Device project at Tokyo University of Science.

\end{document}